# Mid-infrared pulse generation via achromatic quasi-phase-matched OPCPA

B. W. Mayer,[1,*,†] C. R. Phillips,[1,†] L. Gallmann,[1,2] and U. Keller[1]

[1]*Department of Physics, Institute for Quantum Electronics, ETH Zurich, 8093 Zurich, Switzerland*
[2]*Institute of Applied Physics, University of Bern, 3012 Bern, Switzerland*
*[mayerb@phys.ethz.ch](mayerb@phys.ethz.ch)
[†]These authors contributed equally

**Abstract:** We demonstrate a new regime for mid-infrared optical parametric chirped pulse amplification (OPCPA) based on achromatic quasi-phase-matching. Our mid-infrared OPCPA system is based on collinear aperiodically poled lithium niobate (APPLN) pre-amplifiers and a non-collinear PPLN power amplifier. The idler output has a bandwidth of 800 nm centered at 3.4 μm. After compression, we obtain a pulse duration of 44.2 fs and a pulse energy of 21.8 μJ at a repetition rate of 50 kHz. We explain the wide applicability of the non-collinear QPM amplification scheme we used, including how it can enable octave-spanning OPCPA in a single device when combined with an aperiodic QPM grating.

## References and links

## 1. Introduction

High-intensity light sources emitting in the mid-infrared (mid-IR) spectral range have received great attention in recent years for applications including spectroscopy and strong-field physics. As examples, mid-IR sources have been used for spectroscopic investigations of molecular dynamics, and for high harmonic generation (HHG) beyond the water window [1-3]. Probing the wavelength-scaling behavior of strong-field photoionization is another key motivation for developing intense long-wavelength driving fields [4-6]. Ultrafast and high peak-power mid-IR light sources have also been used for accelerator applications, such as enhancing the electron peak current in x-ray free-electron-lasers [7].

Development of laser technologies combining high intensities and few-cycle pulse durations is necessary for many of these applications in order to access the relevant strong-field regimes. Moreover, high repetition rates (and hence high average powers) are important for obtaining high photon fluxes as well as good signal to noise ratios in reasonable time-frames. Optical parametric chirped pulse amplification (OPCPA) is especially attractive in this respect in that it allows few-cycle and high-power laser technologies to be utilized simultaneously (few-cycle for seeding; high-power for pumping). Moreover, new wavelength regions are accessible via OPCPA, provided broadband phase-matching can be achieved.

Several approaches have been explored for broadband phase-matching of OPCPA in the mid-IR. These include work based on periodically poled lithium niobate (PPLN) [8-12], birefringent phase-matching (BPM) [13-15], and aperiodic PPLN (APPLN; also referred to as chirped QPM devices) [16-18]. This latter approach is attractive because it allows broadband phase-matching despite the material's dispersion and damage threshold. For example, assuming 1.064-μm pumping and a near-collinear beam geometry, phase-matching from degeneracy up to ~4.5-μm in $MgO:LiNbO_3$ requires a range of QPM periods from 32.4 μm to 27.5 μm. Since the relative change in period is small, fabricating a grating whose period smoothly varies between these periods is possible with comparable lithographic poling procedures as used for PPLN devices. This property also means that APPLN devices are well suited for mid-IR seed generation via DFG [11, 19].

However, due to a number of design constraints that were explained recently [20], achieving broadband OPCPA supporting high-quality pulses is more challenging than obtaining a sufficient phase-matching bandwidth. In particular, two of the constraints discussed in [20] motivated the present work: First, unwanted nonlinear processes involving sum- and difference-frequency mixing between the nominal pump, signal and idler pulses can be phase-matched due to the broad range of QPM periods involved. Second, when large intensity-times-length products are involved, the device becomes more sensitive to random duty cycle (RDC) errors in the QPM grating [21], which can lead to an unwanted enhancement of pump second-harmonic generation (SHG), even if this process is not phase-matched by the nominal grating design. A new amplification regime, one that is less sensitive to these effects, is therefore called for.

In this paper, we demonstrate OPCPA based on a non-collinear beam configuration in combination with QPM, implemented via PPLN. With this approach, we generate mid-IR pulses at a center wavelength of 3.4 μm with pulse duration 44.2 fs (sub-four-cycles), an energy of 21.8 μJ, and a repetition rate of 50 kHz (average power 1.09 W). This pulse energy represents more than an 80% improvement compared to our most recent result using the same pump and seed lasers. Our system uses APPLN devices for simultaneous pre-amplification and parametric transfer of the 1.5-μm seed to the mid-infrared, combined with a non-collinear PPLN power amplifier operated in an achromatic phase-matching regime, and is the first realization of such a "hybrid" OPCPA configuration. Furthermore, we show the wide applicability of this scheme for broadband parametric amplification and how non-collinear APPLN devices can resolve the issues for aperiodic QPM devices mentioned above, enabling octave-wide gain bandwidths.

## 2. Experimental setup

A schematic of our OPCPA system is shown in Fig. 1. Here, we use a new final power amplifier stage, while keeping the seeding and pre-amplification layout the same as the system reported in [18]. For completeness, we next give an overview of the system layout.

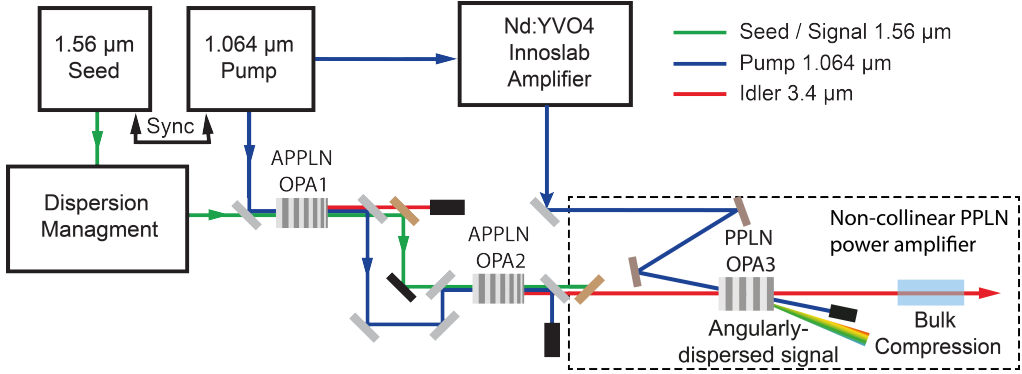

Fig. 1. Experimental setup of the mid-IR OPCPA. The angular-dispersion-free idler extracted from the pre-amplifier OPA2 is used as seed for the non-collinear PPLN power amplifier (OPA3) shown in the dashed box.

The OPCPA system is pumped at 1064-nm by an industrial laser (Time Bandwidth Products "Duetto") which produces 12-ps pulses at a repetition rate of 50 kHz and average power of 10 W. Approximately 6 W from this laser is used for pumping our APPLN-based pre-amplifiers, while the remaining power is used to seed a home-built Innoslab amplifier [22], which currently provides up to 17.8 W when operated at 50 kHz. To seed the OPCPA system, we use a femtosecond fiber laser which produces 65-fs pulses at 82-MHz at a center wavelength of 1560 nm. Theses seed pulses are spectrally broadened in a dispersion-compensated fiber. After this fiber, seed-pulse chirping is performed with a 4-f pulse shaper followed by a silicon prism pair. After the pulse chirping arrangement, there are two collinear pre-amplifiers based on apodized APPLN gratings [23], which we denote as OPA1 and OPA2. As mentioned above, the seeding, chirping, and pre-amplification components of the system are the same as described in [18].

For this work, we modified the power amplifier, denoted OPA3. The new amplifier configuration is depicted in the dashed box of Fig. 1. The collinear pre-amplifier OPA2 is seeded by the 1.5-μm signal output of OPA1, and generates a broadband and angular-chirp-free 3.4-μm idler output suitable for seeding OPA3. OPA3 is based on a PPLN grating and uses a non-collinear pump beam geometry, described further in section 3. Compression of the amplified idler is performed in a 50-mm sapphire rod. The spectral phase of the signal seed is parametrically transferred to the generated idler: by adjusting the signal pulse chirp before OPA1 via the 4-f pulse shaper and Si prism pair, the remaining chirp on the amplified idler pulses after the sapphire rod is compensated.

## 3. Phase-matching and parametric amplification

The phase-matching scheme employed in OPA3 is depicted in Fig. 2. Rather than the conventional *non-critical* configuration enabled by QPM (where phase-matching is insensitive to the incident angles), we use an *achromatic* configuration based on non-collinear beams, in which the phase-mismatch Δk is insensitive to wavelength. In this configuration, QPM offers two unique advantages over conventional birefringent phase-matching (BPM): the ability to achieve phase-matching for almost any beam geometry while still using the largest nonlinear coefficient of a material; and the ability to use aperiodic QPM devices for even larger gain bandwidths and customized gain spectra.

The corresponding phase-matching diagram is shown in Fig 2(a), assuming mid-IR idler seeding collinear to the constant grating k-vector $K_g$, and a narrow-band non-collinear pump incident at an angle to the crystal c-axis. The inset shows the frequency-dependent spread of the longitudinal phase-mismatch Δk by zooming in on the circled region. For comparison, Fig. 2(b) shows the phase-matching diagram for two spectral components (3 μm and 4 μm) in our collinear APPLN pre-amplifiers (the scale and changes in $K_g$ are exaggerated). The schematic of a chirped QPM grating shown next to these phase-matching diagrams illustrates how these spectral components are amplified around different positions in the grating (3 μm near the lower, input-side of the grating; 4 μm near the upper, output-side).

In Fig. 2(c), we show the corresponding frequency-dependent phase-mismatch in the absence of a QPM grating, denoted $\Delta k_0(\lambda)$, for both collinear and non-collinear cases. The non-collinear case (green curve) shows a relatively small spread of $\Delta k_0(\lambda)$: in this case, a single QPM period in a short device combined with an intense pump can amplify the 3- to 4-μm spectral range, as in OPA3. In the collinear case (blue curve), the wider spread of $\Delta k_0(\lambda)$ can still be amplified by a longer, chirped grating having the corresponding range of QPM periods, as in our pre-amplifiers (see dot-dashed lines around the blue curve). The potential of combining a chirped QPM grating with the non-collinear arrangement is illustrated by the dot-dashed lines around the green curve: in this case, an octave-spanning gain spectrum is supported using a similar spread of grating periods as we use in our current pre-amplifiers OPA1 and OPA2. We discuss this possibility, and the importance and constraints of combining chirped QPM gratings with non-collinear geometries, in more detail in section 5.

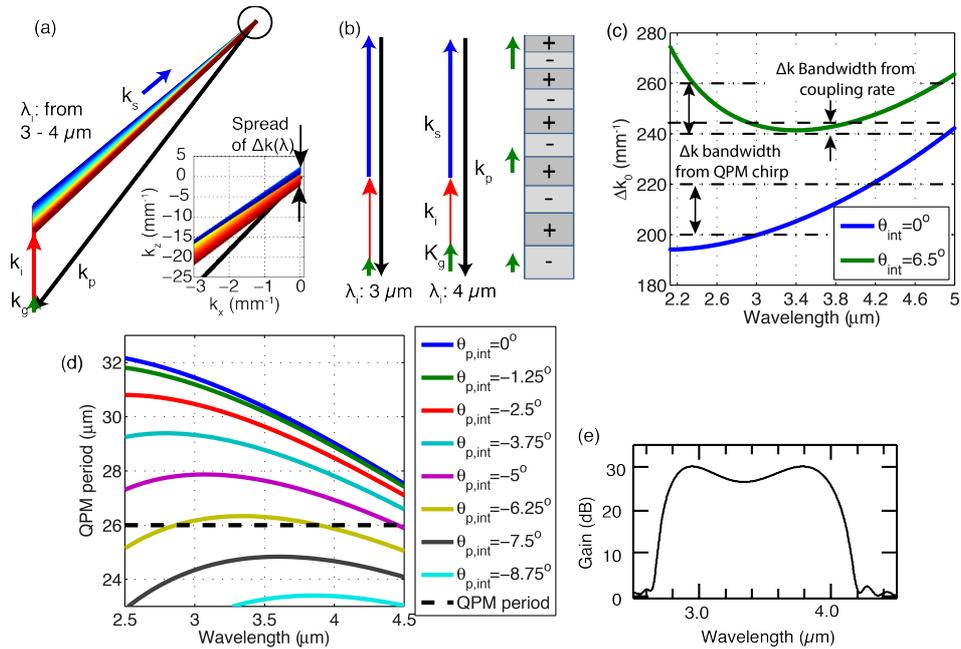

Fig. 2. (a): Illustration of the achromatic phase-matching scheme used for the power amplifier (OPA3). The frequency-dependent signal angle is chosen to yield zero phase mismatch in the transverse direction. Inset: zoom-in on the circled region, showing the associated spread of $\Delta k(\lambda)$ in the longitudinal direction across the pulse spectrum. (b) Collinear phase-matching diagram for two spectral components in a chirped QPM device (e.g. OPA1 and OPA2), for comparison to the non-collinear scheme. The chirped grating structure is shown along with three green arrows indicating the spatially-varying grating k-vector $K_g$. (c) Phase-mismatch $\Delta k_0(\lambda)$ in the absence of a QPM grating for the non-collinear (green curve) and collinear (blue curve) configurations shown in (a) and (b), respectively. (d): Phase-matching period for several different angles of the pump beam. The QPM poling-period of 26 μm used for OPA3 is indicated by the black dashed line. (e) Example small-signal gain spectrum enabled by such a device.

To illustrate the role of the pump angle in more detail, in Fig. 2(d) we plot the frequency-dependent quasi-phase-matching period as a function of wavelength for several internal angles of the pump beam. Generally, the broadest bandwidth from a PPLN device is obtained near the turning point in QPM period with respect to wavelength; this turning minimizes the spread of Δk [Fig. 2(a) inset], and also corresponds to group velocity (GV) matching of the signal and idler waves. By varying the pump angle (see legend), the turning point can be tuned to any wavelength beyond degeneracy. In the experiment, we use an internal pump angle of $\Theta_{int} \sim 6.25°$, and a QPM period of 26 μm, as indicated by the dashed black line in Fig. 2(d).

Figure 2 shows how the beam geometry can be optimized for bandwidth [minimizing the spread of Δk(λ)], and phase-matching can be obtained at whatever angle is optimum by adjusting the QPM period. The analogous procedure in BPM is to rotate the crystal axes, but in this case both the beam geometries and the nonlinear tensor elements that can be used are more strongly constrained by phase-matching considerations, and there is no straightforward way of introducing a spatially varying phase mismatch.

Similar phase-matching schemes to the one discussed above have been deployed previously. Non-collinear QPM in PPLN was demonstrated in the context of 1.064-μm-pumped optical parametric generation (OPG) of a broadband signal wave in the spectral range between 1.66 μm and 1.96 μm [24]. A related non-collinear phase-matching scheme, based on an angularly chirped input signal wave, was proposed for PPLN in Ref. [25]. Until now, however, such schemes have not been experimentally demonstrated for broadband QPM OPCPA. Furthermore, beyond our experimental demonstration (which is described in section 4), we explain in section 5 the opportunities enabled by non-collinear pumping of aperiodic QPM devices.

We also note that small non-collinearities are typically employed in wavelength-degenerate OPA and OPCPA schemes for beam separation and to avoid the phase-sensitivity of true degenerate OPA; example systems using QPM include Refs. [8-11]. The small non-collinearity in such systems only has a minor influence on the phase-matching properties, unlike the approach we use here.

Next, to illustrate the characteristic features of OPCPA based on the phase-matching curves shown in Fig. 2(c) and 2(d), an example gain spectrum for a mid-IR PPLN OPA device is shown in Fig. 2(e). For this example we assume a moderate coupling rate between the signal and idler waves of $\gamma=2$ mm$^{-1}$. The corresponding small-signal gain spectrum is given by

$$G_{ss} = \left| \cosh(gL) + i\frac{\Delta k}{2g}\sinh(gL) \right|^2 \quad (1)$$

where $g=[\gamma^2 - (\Delta k/2)^2]^{1/2}$ and L is the crystal length. A finite Δk at the center wavelength can broaden the gain spectrum, as for collinear and nearly-collinear degenerate OPA [11]. However, that the phase mismatch should not be too large for a saturated amplifier, because the achievable pump depletion is also limited by the ratio (Δk/γ). Assuming only two of the three waves are present at the input of the device (i.e. only the pump and the seed waves), and taking the high-gain OPA limit of such parametric interactions (i.e. by assuming a very weak input seed wave compared to the pump but a long enough crystal to fully deplete the pump), the maximum plane-wave conversion efficiency is given by

$$\eta_{max} = 1 - \left(\frac{\Delta k}{2\gamma}\right)^2 \quad (2)$$

where this equation applies for |Δk/2γ|<1 (which is the requirement for parametric gain). This result can be obtained via the formalism described in [26], which provides an intuitive geometrical interpretation of the possible parametric conversion trajectories given arbitrary input conditions and Δk. A value of Δk/γ=-0.6 at the center wavelength thus provides a reasonable trade-off: the bandwidth is broadened according to Eq. (1) and Fig. 2(e), while the maximum efficiency (for an idealized plane-wave interaction) is still >90% based on Eq. (2).

## 4. Experimental Results

We implemented the proposed scheme as the final power amplification stage. The average powers for the seed and pump just before the anti-reflection (AR) coated 2-mm long, 2-mm by 2-mm wide PPLN crystal are 31 mW and 17.8 W, respectively. The pump is focused to an intensity of 10.7 GW/cm$^2$ (1/e$^2$ radius of 475 μm in the horizontal and 350 μm in the vertical axis). In Fig. 3(a), we show the measured idler spectra before and after the power amplifier.

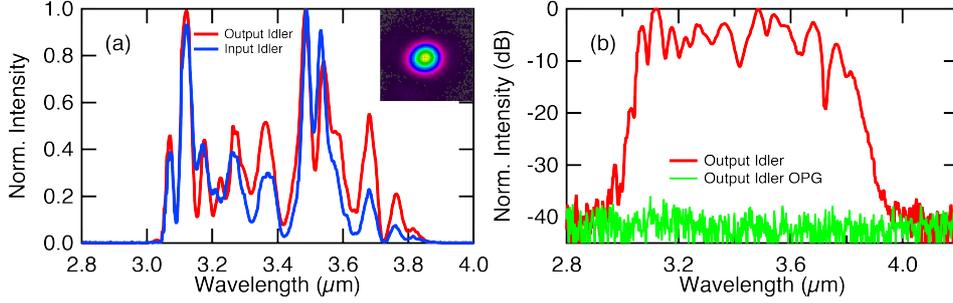

Fig. 3. (a): Measured spectra of the input idler (blue) and output idler (red) of the non-collinear OPCPA power amplifier. The inset shows the good input idler beam profile produced by the second APPLN pre-amplifier, OPA2. (b) Spectral characterization of the OPG background, measured by blocking the near-IR seed pulses before the first APPLN pre-amplifier, in comparison to the idler pulse spectrum. Our measurements show that the energy of this OPG background is at least 34.6 dB lower than the idler pulse energy.

Next, to estimate the optical parametric generation (OPG) background, the signal seed was blocked before the first APPLN pre-amplifier. The spectrally resolved OPG background is shown in Fig. 3(b). We also measured its average power with a thermal power meter. The spectral characterization as well as the power measurements are within the noise floor of the respective detectors, indicating a negligible OPG background compared to the amplified pulses (<300 μW). Furthermore, no evidence for photorefractive effects was observed.

The amplified idler pulses were compressed by the careful design of sign and amount of dispersion of the near-IR seed pulses prior to the APPLN pre-amplifiers, in combination with bulk-propagation of the idler after the final power amplifier through an AR-coated, 50-mm-long sapphire rod. This method yielded a pulse duration of 44.2 fs (less than four optical cycles). Figure 4 shows the measured and retrieved spectrograms, spectra and spectral phase from an SHG-FROG (Frequency Resolved Optical Gating) characterization of the pulses. The residual fluctuation of the spectral phase results in a pedestal in the time domain. Note that, as discussed in [18], the fluctuations in spectral phase originate from our near-IR seeding scheme, rather than any of the OPCPA devices. The reconstructed temporal intensity profile is shown in Fig. 4(d). The energy of the compressed idler pulse was 21.8 μJ, corresponding to an average power of 1.09 W at the 50-kHz repetition-rate and a gain of 15.46 dB.

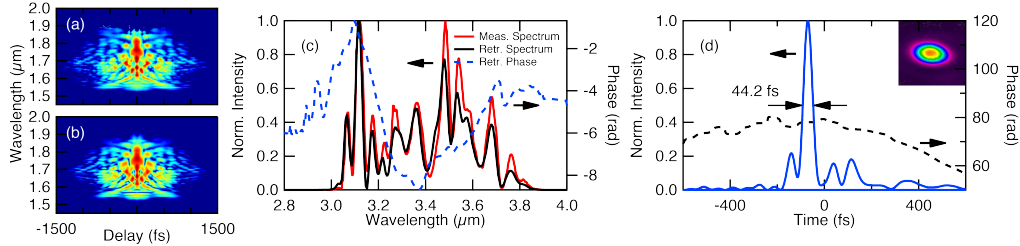

Fig. 4. (a): Measured SHG-FROG spectrogram. (b): Retrieved spectrogram. (c): Measured (red) and retrieved (black) idler spectrum and spectral phase (blue dashed) after compression. (d): Retrieved temporal intensity profile. The inset shows the idler beam profile. The FROG-error of the reconstruction for a grid size of 512 by 512 points is 0.0125.

The inset of Fig. 4(d) shows the beam profile of the compressed mid-IR idler. The fairly large non-collinear angle between the idler, signal and pump beams leads to some spatial walk-off within the crystal. This effect, combined with the slight ellipticity of our pump beam, explains the slight ellipticity of the idler beam. Moreover, the spatial walk-off effect imposes a constraint on the peak power of the pump that is needed to obtain a high parametric gain (while using a round beam). For a characteristic non-collinear angle θ, the ratio of transverse spatial walk-off $w_{wo}$ to pump beam radius $w_p$ can be expressed as

$$\frac{w_{wo}}{w_p} = \frac{(\gamma L)\tan(\theta)}{(\gamma w_p)} \approx (P_0/P_{p,pk})^{1/2} \operatorname{arcosh}(\sqrt{G_{ss}})\tan(\theta) \quad (3)$$

In the first part of this equation, the value of θ is determined by the group velocity matching condition; the (γL) factor determines the small-signal OPA gain [27]; and, for a given nonlinear crystal, $(\gamma w_p)^2$ is proportional to the peak power of the pump. The second part of the equation applies these relations assuming Δk=0 for calculating $G_{ss}$; $P_0$ is a characteristic peak power associated with the material and the wavelengths involved, and is given by $P_0 = (n_i n_s n_p \varepsilon_0 c \lambda_s \lambda_i)/(16\pi d_{eff}^2)$.

For a 50-% duty cycle QPM grating and $d_{33}$=19.5 pm/V for MgO:LiNbO$_3$ [28], $P_0$=17.2 MW. Consequently, with θ=6.25° (the geometry we use experimentally), a peak power of 14 MW is required to achieve a plane-wave small-signal gain of 30-dB while keeping $w_{wo}/w_p$ < 0.5 (to avoid large walk-off relative to the pump beam size). The peak pump power we use in OPA3 is 24.1 MW, satisfying this constraint. To satisfy the walk-off constraint with much lower peak powers, an elliptical pump beam would be required.

## 5. Implications for other QPM media

### 5.1. Aperiodic QPM devices

Looking beyond OPCPA in periodic devices, our results open up the interesting possibility of combining achromatic quasi-phase-matching arising from non-collinear beams with aperiodic (chirped) QPM gratings. In this section, we show the advantages of this approach, and discuss the peak power levels required.

In section 6 of [20], we considered the possibility of phase-matching additional sum and different frequency mixing processes in chirped QPM devices. If the quasi-phase-matching period for some unwanted processes (such as idler SHG) overlaps with the QPM periods required for the nominal OPA process, then this unwanted process may be efficient, potentially distorting the OPCPA output. Experimentally, we avoided these effects by the choice of QPM grating chirp rate [18, 20], but as the range of QPM periods is increased (to amplify a broader bandwidth), it becomes more difficult to avoid such processes in a collinear geometry.

However, if the desired OPA process depends on the beam angles in a different way to the other DFG and SFG processes, then the beam geometry offers a way to reduce the influence of these other effects. To illustrate this approach, Fig. 5 shows the phase-matching period for several processes assuming different pump angles and wavelengths: for 0° and $\lambda_p$=1064 nm (a), and for 7.5° and $\lambda_p$=1030 nm (b); the legend applies to both (a) and (b).

As an example, consider the OPA and idler SHG processes. To suppress idler SHG, this process should be phase-matched before the OPA process (or it should not be phase-matched at all). However, if idler SHG occurs *later* in the grating than OPA, the idler will be intense (especially in an OPA stage with high pump depletion) and so idler SHG will be efficient. In (a), the curves for OPA and idler SHG cross, so the latter process cannot be avoided for all the wavelengths shown. In (b), these processes do not cross and the required QPM periods do not overlap. Similar conclusions can be drawn from inspection of the other processes shown: thus, most of the parasitic processes can be avoided in (b) by using a QPM period that increases with position through the grating. There is an exception at 2.6 μm where the curves for OPA and signal SHG cross, but the signal wave is discarded.

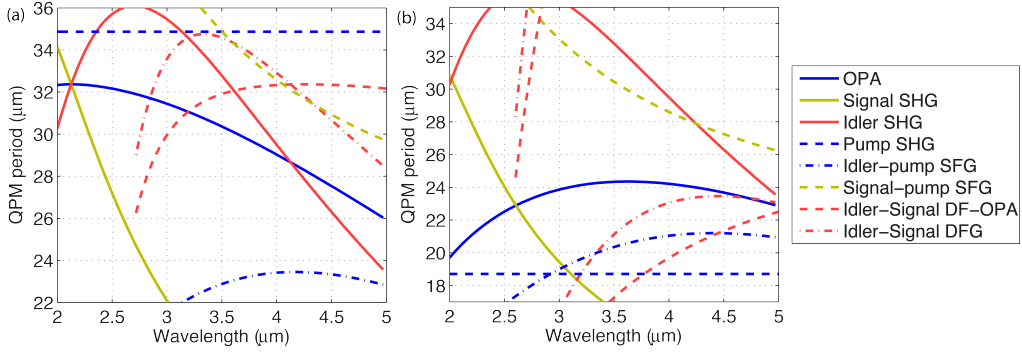

Fig. 5. Phase-mismatch for unwanted processes for two pump wavelengths and angles geometries. (a) $\theta_p=0°$ and $\lambda_p=1.064$ nm, i.e. collinear beams, as currently used in our APPLN devices and discussed in [20]. (b) Non-collinear configuration with $\theta_p=7.5°$ and $\lambda_p=1030$ nm. We assume idler seeding with angle 0° to the grating k-vector. The plane of incidence includes the crystal c-axis. Several processes are shown: SHG of the nominal pump, signal and idler waves; SFG of the idler and signal with the pump; DFG between the signal and idler waves; and amplification of this signal-idler difference frequency by the pump. The signal is discarded, so processes involving this wave are less critical. The curves are cut off when one of the wavelengths involved lies beyond 5 µm, based on the high absorption and inaccuracy of the Sellmeier relation beyond this wavelength [29]. In (a), the idler SHG process cannot be avoided across the whole wavelength range shown, as discussed in the text. In (b), with a positive QPM chirp rate, corresponding to a QPM period increasing with position through the grating from 21.6 µm to 24.4 µm, idler wavelengths from 2.3 µm to 5 µm (and possibly further, limited by absorption and accuracy of the Sellmeier equation) can be amplified while suppressing the parasitic processes shown. .

Figure 5 shows how the OPCPA process can be favored over other parasitic processes such as idler SHG. The amplification bandwidth supported is as much as 2.3 µm – 5 µm, more than an octave. The corresponding range of grating k-vectors is 258 mm$^{-1}$ – 291 mm$^{-1}$ (QPM periods from 24.4 µm to 21.6 µm). Since this grating spatial frequency bandwidth is only slightly larger than that used in our current OPCPA experiments, the device should not be much more sensitive to the effects of RDC errors [20, 21].

Nonetheless, a potential issue with this non-collinear APPLN OPCPA scheme is spatial walk-off: if there is an idler-pump beam walk-off comparable to their beam sizes, then idler spectral components phase-matched near the start of the chirped QPM grating will be displaced laterally compared to those phase-matched near the end, leading to a spatial chirp. This spatial walk-off must therefore be minimized. The relative amount of walk-off can be expressed in terms of the small-signal gain, in analogy to Eq. (3). Noting that the small-gain is given, for a linearly chirped QPM grating, by $G_{ss}=\exp(2\pi\Lambda)$ [30], where $\Lambda=\gamma^2/|\Delta k'|$ and $\Delta k'=d(\Delta k)/dz=-dK_g/dz$ is the QPM chirp rate, we arrive at the following:

$$\frac{w_{wo}}{w_p} \approx \left(\frac{|\Delta k'|L}{\gamma}\right)\left(\frac{P_0}{P_{p,pk}}\right)^{1/2}\frac{\ln(G_{ss})\tan(\theta)}{2\pi}, \qquad (4)$$

where $P_0$ was defined following Eq. (3). With $\gamma=3$ mm$^{-1}$ (peak intensity of approximately 10 GW/cm$^2$), $G_{ss}=30$ dB, $\theta=7.5°$, and $|\Delta k'|L=33$ mm$^{-1}$ based on the k-space bandwidth requirement stated above, the corresponding peak power requirement to satisfy $w_{wo}/w_p<0.2$ (imposing a stricter constraint than above because of the greater severity of a spatial chirp effect compared to an elliptical beam distortion) is 1 GW. For a given phase-matching bandwidth, this peak power can be reduced by increasing $\gamma$, but the maximum possible value of $\gamma$ is limited by the damage threshold. Our current pump laser has lower power (~25 MW), but GW peak powers can be achieved with current 1-µm laser technologies. Demonstration of this regime may therefore require an upgraded pump, or alternatively elliptical input beams.

Maintaining small beam walk-off for a relatively large non-collinear angle should also ensure suppression of gain guided non-collinear modes [20, 31], although some care should be taken in this respect, especially if using elliptical beams and lower peak powers than those suggested by Eq. (4). In particular, using a large beam size in the walk-off direction (major axis) but a small one in the other transverse direction (minor axis) increases the susceptibility to gain guided modes along the minor axis. A detailed study of this aspect of the problem is beyond the scope of this paper

*5.2. Orientation-patterned gallium arsenide*

It is interesting to note that the amplification technique we have demonstrated is applicable to other QPM media as well, including in non-birefringent materials. For example, it could support broadband OPA in orientation-patterned GaAs (OP-GaAs) [32], with the potential for extending OPA and OPCPA into the far-IR spectral region [33].

To compare OP-GaAs to MgO:PPLN, note that our 1.064-µm pump wavelength is somewhat longer than half the zero-dispersion wavelength (which is around 1.9 µm in MgO:PPLN). An analogous configuration for OPA in OP-GaAs, whose zero-dispersion wavelength is around 6.6 µm [34], would be to use a pump wavelength of ~3.4 µm or longer. This pumping configuration could support group velocity (GV) matched OPCPA for center seed wavelengths from 7 to ~18 µm. Alternatively, using 2-µm or 2.5-µm pump lasers, GV-matching could be achieved for signal-seeded OP-GaAs devices. Therefore, the technique we have demonstrated may prove to be important in developing widely tunable and few-cycle sources in the far-IR as well as the mid-IR.

**6. Conclusion and outlook**

In conclusion we have demonstrated a broadband mid-IR hybrid OPCPA system based on APPLN pre-amplification and a new non-collinear group-velocity-matched PPLN power amplifier. The idler seed pulses, centered around 3.4 µm, are generated from two weakly-saturated all-collinear APPLN pre-amplifiers. The pre-amplifiers take advantage of important features of chirped QPM media, including convenient and collinear parametric transfer of the infrared (1.5 µm) seed to the mid-IR, and amplification of a broad bandwidth determined by the QPM grating structure. Such pre-amplifiers are also less sensitive to some of the parasitic processes shown in Fig. 5(a), since the signal and idler waves remain significantly weaker than the pump throughout the device.

The idler pulses generated by OPA2 are used to seed OPA3, where they are amplified to 21.8-µJ (energy measured after pulse compression), representing an 80% improvement over our previous result. The pulses are compressed to 44.2 fs (corresponding to less than four optical cycles) by propagation through a bulk sapphire rod. Fine-tuning of the idler chirp is obtained by adjusting the pre-chirping of the 1.56-µm signal seed before the pre-amplifiers.

The group-velocity-matched non-collinear QPM OPCPA technique demonstrated here is a highly versatile approach for broadband amplification in the mid-infrared, and even the far-infrared via OP-GaAs. Our results also demonstrate that this technique can be combined with APPLN amplifiers in a practical "hybrid" OPCPA system configuration. Furthermore, we showed in section 5.1 how the combination of non-collinear beam geometry and chirped QPM devices enables octave-spanning amplification; the use of nonlinear chirp profiles would even enable customizable gain profiles across this amplification bandwidth [30, 35]. This approach, together with further power scaling achievable by the use of wide-aperture QPM devices as the gain medium [36], thus provides a route to amplification of high-energy single-cycle pulses. This capability will enable a wide variety of attosecond science experiments across the mid-infrared wavelength region.

**Acknowledgments**

This research was supported by the Swiss National Science Foundation (SNSF) through grant #200020_144365/1, and by a Marie Curie International Incoming Fellowship within the 7th European Community Framework Programme.